\documentclass[11pt]{article}
\def\appendix{\par\clearpage
  \setcounter{section}{0}
  \setcounter{subsection}{0}
  \@addtoreset{equation}{section}
  \def\@sectname{Appendix~}
  \def\theequation{\thesection\arabic{equation}}
  \def\thesection{\Alph{section}}}

\def\thefigures#1{\par\clearpage\section*{Figures\@mkboth
  {FIGURES}{FIGURES}}\list
  {Fig.~\arabic{enumi}.}{\labelwidth\parindent\advance
\labelwidth -\labelsep
      \leftmargin\parindent\usecounter{enumi}}}

\def\thetables#1{\par\clearpage\section*{Tables\@mkboth
  {TABLES}{TABLES}}\list
  {Table~\Roman{enumi}.}{\labelwidth-\labelsep
      \leftmargin0pt\usecounter{enumi}}}

\def\@sect#1#2#3#4#5#6[#7]#8{\ifnum #2>\c@secnumdepth
     \def\@svsec{}\else
     \refstepcounter{#1}\edef\@svsec{\@sectname\csname the#1\endcsname
.\hskip 1em }\fi
     \@tempskipa #5\relax
      \ifdim \@tempskipa>\z@
        \begingroup #6\relax
          \@hangfrom{\hskip #3\relax\@svsec}{\interlinepenalty \@M #8\par}
        \endgroup
       \csname #1mark\endcsname{#7}\addcontentsline
         {toc}{#1}{\ifnum #2>\c@secnumdepth \else
                      \protect\numberline{\csname the#1\endcsname}\fi
                    #7}\else
        \def\@svse=chd{#6\hskip #3\@svsec #8\csname #1mark\endcsname
                      {#7}\addcontentsline
                           {toc}{#1}{\ifnum #2>\c@secnumdepth \else
                             \protect\numberline{\csname the#1\endcsname}\fi
                       #7}}\fi
     \@xsect{#5}}

\def\@sectname{}
\newcommand{\nc}{\newcommand}
\nc{\spa}[3]{\left\langle#1\,#3\right\rangle}
\nc{\spb}[3]{\left[#1\,#3\right]}
\nc{\ksl}{\not{\hbox{\kern-2.3pt $k$}}}
\nc{\hf}{\textstyle{1\over2}}
\nc{\pol}{\varepsilon}
\nc{\tq}{{\tilde q}}
\nc{\esl}{\not{\hbox{\kern-2.3pt $\pol$}}}

\newcommand{\be}{\begin{equation}}
\newcommand{\ee}{\end{equation}\noindent}
\newcommand{\bear}{\begin{eqnarray}}
\newcommand{\ear}{\end{eqnarray}\noindent}
\newcommand{\benn}{\begin{enumerate}}
\newcommand{\enn}{\end{enumerate}}

\date{}
\renewcommand{\theequation}{\arabic{section}.\arabic{equation}}
\renewcommand{\arraystretch}{2.5}

\renewcommand{\arraystretch}{2.5}
\newcommand{\slD}{\raise.15ex\hbox{$/$}\kern-.57em\hbox{$D$}}
\newcommand{\slpartial}{\raise.15ex\hbox{$/$}\kern-.57em\hbox{$\partial$}}
\newcommand{\slG}{{{\dot G}\!\!\!\! \raise.15ex\hbox {/}}}

\def\GBd12{{\dot G}_{B12}}

\def\non{\nonumber}
\def\beqn*{\begin{eqnarray*}}
\def\eqn*{\end{eqnarray*}}

\def\square{\kern1pt\vbox{\hrule height 1.2pt\hbox{\vrule width 1.2pt
   \hskip 3pt\vbox{\vskip 6pt}\hskip 3pt\vrule width 0.6pt}
   \hrule height 0.6pt}\kern1pt}

\def\slash#1{#1\!\!\!\raise.15ex\hbox {/}}

\def\dps{\displaystyle}

\def\half{{1\over 2}}

\def\fourth{{1\over4}}

\def\e{\mbox{e}}

\def\4piTD{{(4\pi T)}^{-{D\over 2}}}
\def\4piT4{{(4\pi T)}^{-2}}

\def\Tintm4{{\dps\int_{0}^{\infty}}{dT\over T}\,e^{-m^2T}
    {(4\pi T)}^{-2}}
\def\Tintm{{\dps\int_{0}^{\infty}}{dT\over T}\,e^{-m^2T}}

\def\arraystretch{2.5}

\def\mn{{\mu\nu}}

\def\nc#1#2#3{     {\it Nuovo Cim. }{\bf #1} (19#2) #3}

\def\bbbz{{\mathchoice {\hbox{$\sf\textstyle Z\kern-0.4em Z$}}
{\hbox{$\sf\textstyle Z\kern-0.4em Z$}}
{\hbox{$\sf\scriptstyle Z\kern-0.3em Z$}}
{\hbox{$\sf\scriptscriptstyle Z\kern-0.2em Z$}}}}

\begin{document}
\pagestyle{empty}
\renewcommand{\thefootnote}{\fnsymbol{footnote}}
\hfill {\sl UMSNH-Phys/02-13}
\vskip .4cm
\begin{center}
{\huge\bf On the low-energy limit of the}\\
{\huge\bf  QED N -- photon amplitudes}\\
\vskip1.3cm
{\large Louise C. Martin$^{a}$, Christian Schubert$^{a,b}$,
Victor M. Villanueva Sandoval$^{a}$}
\\[1.5ex]

\begin{itemize}
\item [$^a$]
{\it
Instituto de F\'{\i}sica y Matem\'aticas
\\
Universidad Michoacana de San Nicol\'as de Hidalgo\\
Edificio C-3, Apdo. Postal 2-82\\
C.P. 58040, Morelia, Michoac\'an, M\'exico\\
}
\item [$^b$]
{\it
California Institute for Physics and Astrophysics\\
366 Cambridge Ave., Palo Alto, CA 94306, USA}
\end{itemize}
\vskip 2cm
\vspace{20pt}
 {\large \bf Abstract}
\end{center}
\begin{quotation}
\noindent
We derive an explicit formula for the low energy limits of the
one-loop on-shell massive $N$ - photon amplitudes, for arbitrary $N$ and
all helicity assignments, in scalar and spinor QED.
The two-loop corrections to the same amplitudes are obtained
for up to the ten point case.
All photon amplitudes with an odd number of
`+' helicities are shown to vanish in this limit
to all loop orders.

\end{quotation}
\vskip 1cm
\clearpage
\renewcommand{\thefootnote}{\protect\arabic{footnote}}
\pagestyle{plain}

\section{Introduction: On-shell QED photon amplitudes}
\label{introduction}
\renewcommand{\theequation}{1.\arabic{equation}}
\setcounter{equation}{0}

In recent years substantial progress has been made in the
computation of on-shell one-loop amplitudes.
This has been due to
the development of new techniques \cite{berkos,bernreview}
which provide alternatives
to the standard Feynman diagrammatic approach, as well as
to progress in the calculation of the basic integrals
\cite{davyd,bedikopent,caglmi,pittau,weinzierl,fljeta,biguhe,bghs}.
Much of this work has been concerned with massless amplitudes,
which are computationally the most accessible ones.
It led to a number of unexpectedly simple results for certain special
helicity configurations of photon or gluon amplitudes
(see \cite{bernreview} for a review).
A particularly striking result is Mahlon's observation that the
massless one-loop QED $N$ - photon amplitudes with all helicities
equal vanish on-shell for
all $N\geq 6$ \cite{mahlon}.

For the corresponding amplitudes involving massive
loops little is known beyond the four-point case
\cite{karneu,cotopi,pasvel,denner}.
In the present paper, we will investigate the QED
$N$ photon amplitudes in the limit of low photon energies, i.e.
with photon momenta such that all kinematic invariants
$k_i\cdot k_j$ are small compared to $m^2$
(see \cite{dicare} for a discussion of this approximation for
the photon scattering case). As is well-known
(see, e.g., \cite{itzzubbook}), in this limit the photon amplitudes
are directly related to the QED effective Lagrangian
${\cal L}(F)$ for a background
field with a constant field strength tensor $F_{\mu\nu}$. Namely,
to obtain the amplitude with photon momenta
$k_1,\ldots ,k_N$ and polarisation vectors
$\pol_1,\ldots ,\pol_N$, define for every leg the
field strength tensor

\bear
F_i^{\mu\nu} &\equiv & k_i^{\mu}\pol_i^{\nu} -  k_i^{\nu}\pol_i^{\mu}
\ear
and

\bear
F_{\rm tot} &\equiv & \sum_{i=1}^N F_i
\label{defFtot}
\ear
The corresponding amplitude is then obtained by inserting
$F_{\rm tot}$ into the effective Lagrangian, and extracting
the terms involving each $F_1,\ldots,F_N$ precisely once:

\bear
\Gamma^{(EH)}
[k_1,\varepsilon_1;\ldots;k_N,\varepsilon_N]
&=&
{\cal L}\bigl(iF_{\rm tot}\bigr)\bigg\vert_{F_1\cdots F_N}
\label{defEHamp}
\ear
At one loop, the QED effective Lagrangian for the constant field strength
case is just the
well-known Euler-Heisenberg
Lagrangian \cite{eulhei,weisskopf} whose
weak field expansion is known in closed form.
Nevertheless, it appears that (\ref{defEHamp}) was previously applied
only to the textbook case of photon-photon scattering.

In \cite{dunsch2,dunsch3} G.V. Dunne and one of the authors had
considered the special case of constant {\it self-dual} background
fields, and
derived closed-form expressions for the corresponding
two-loop Euler-Heisenberg Lagrangian and its scalar QED analogue.
Due to the well-known correspondence between self-dual fields
and helicity eigenstates
\cite{dufish,bardeen,selivanov,cangemi,chasie}
this Lagrangian still contains the full information on the
low energy limit of the all `+' component of the $N$ - photon
amplitudes in the helicity decomposition. The relation
(\ref{defEHamp}) could thus be used to derive simple closed-form
expressions for these amplitudes not only at one but also at
two loops, for arbitrary $N$, in scalar and spinor QED.

In the present paper we extend the same approach to
the case of arbitrary helicity assignments.
Applying (\ref{defEHamp})
to the Euler-Heisenberg Lagrangian
and its scalar QED analogue will allow us to
obtain closed-form
expressions for the low energy limits of the one-loop $N$-photon
amplitudes with arbitrary helicity assignments.
The standard spinor helicity technique \cite{bkdgw,klesti,xuzhch}
will
turn out highly useful in working out the algebra of the
field strength tensors $F_i$.

Although various integral representations have been derived
for the corresponding two-loop effective Lagrangians
\cite{ritusspin,ritusscal,ginzburg,ditreuqed,rss,frss,korsch},
for the case of a general constant field none of them
is sufficiently explicit to obtain corresponding all - $N$ formulas
at the two - loop level. Nevertheless, we will use the formulas
given in \cite{ritusspin,ritusscal,ginzburg}
to obtain the weak-field expansions of these
two -- loop effective Lagrangians up to the order $(F^{10})$, which will
allow us to compute the corresponding photon amplitudes up to
the ten point case.

\section{One loop spinor QED}
\label{section1lspin}
\renewcommand{\theequation}{2.\arabic{equation}}
\setcounter{equation}{0}

Let us begin with spinor QED at the one-loop level.
We will use the standard integral representation
of the Euler-Heisenberg Lagrangian \cite{eulhei},

\bear
{\cal L}_{\rm spin}^{(1)}
&=&
-
{1\over 8\pi^2}
\int_0^{\infty}{dT\over T}
\,\e^{-m^2T}
\biggl\lbrack
{e^2ab\over \tanh(eaT)\tan(ebT)}
-{e^2\over 3}\bigl(a^2-b^2\bigr)
-{1\over T^2}
\biggr\rbrack
\non\\
\label{L1spin}
\ear
Here $T$ denotes the propertime of the
loop fermion, and $a,b$ are related to the two
invariants of the Maxwell field by
$a^2-b^2=B^2-E^2, ab = {\bf E}\cdot{\bf B}$.
The charge $e$ will often be set to unity in the following.
The subtraction of the terms of zeroeth and second order
in $a,b$ corresponds to on-shell renormalization.
These terms are not relevant for our purposes and
will be omitted in the following.
The invariants
$a,b$ can be related to the field strength tensor $F_{\mu\nu}$
and its dual
\footnote{We work in Minkowski space with
$\eta = {\rm diag}(1,-1,-1,-1)$ and $\varepsilon_{0123}=1$.}
$\tilde F_{\mu\nu} =
\half\varepsilon_{\mu\nu\alpha\beta}F^{\alpha\beta}$,

\bear
a^2 &=&
\fourth \sqrt{\bigl(F_{\mu\nu}F^{\mu\nu}\bigr)^2
+\bigl(F_{\mu\nu}\tilde F^{\mu\nu}\bigr)^2}
+\fourth F_{\mu\nu}F^{\mu\nu}
\non\\
b^2 &=&
\fourth \sqrt{\bigl(F_{\mu\nu}F^{\mu\nu}\bigr)^2
+\bigl(F_{\mu\nu}\tilde F^{\mu\nu}\bigr)^2}
-\fourth F_{\mu\nu}F^{\mu\nu}
\non\\
\label{relabFFtilde}
\ear

We wish to use this Lagrangian to obtain the low energy
limit of the
on-shell $N$ - photon amplitude for arbitrary
$N\geq 4$ and with an arbitrary
helicity assignment. Due to Furry's theorem we can, of course,
assume that $N$ is even.
Since in the abelian case the ordering of the legs does not
matter we shall further assume that legs $1,\ldots, K$ carry
the helicity `+' and the remaining ones helicity `-'.
Also, by CP invariance flipping all helicities
is equivalent to changing all momenta from outgoing to
ingoing. It is therefore sufficient to consider the
case $K\geq N-K$.
To construct suitable polarisation vectors we use the
standard spinor helicity formalism.
In this formalism, a polarisation vector with circular polarisation
`$\pm$' for a photon with momentum $k$ is written as
\bear
\varepsilon^{\pm}_{\mu} &=&
\pm
{\langle q^{\mp}\mid\gamma_{\mu}\mid k^{\mp}\rangle
\over
\sqrt{2}\langle q^{\mp}\mid k^{\pm}\rangle}
\label{eps+-}
\ear
Here
$\langle q^{\pm}\mid k^{\mp}\rangle = \overline{u_{\pm}(q)}u_{\mp}(k)$
etc. are basic spinor products, and $q$ is a reference momentum
(see \cite{dixon} for details and conventions).
Changes of the reference momentum amount to gauge transformations.
As usual we will use the notation

\bear
\langle ij\rangle
&\equiv& \langle k_i^-\mid k_j^+\rangle
\non\\
\lbrack ij\rbrack
&\equiv& \langle k_i^+\mid k_j^-\rangle
\non\\
\label{defbrackets}
\ear

So, let us use (\ref{eps+-}) with some arbitrary choice of
reference momenta $q_i$ to define polarisation vectors
$\pol_1^+,\ldots,\pol_K^+,\pol_{K+1}^-,\ldots,\pol_N^-$.
In the corresponding field strength tensor for leg $i$

\bear
F_{i\mu\nu}^{\pm} &\equiv & k_{i\mu}\pol_{i\nu}^{\pm} - \pol_{i\mu}^{\pm}k_{i\nu}
\label{defFipm}
\ear
the dependence on $q_i$ already drops out, as is easily verified.
Using standard manipulations (see, e.g., \cite{dixon}) the
following identities are found to hold

\bear
\lbrace F^+_i,F^+_j \rbrace^{\mu\nu} &=& -\half [ij]^2\eta^{\mu\nu}
\label{anticom+}\\
\lbrace F^-_i,F^-_j\rbrace^{\mu\nu}
&=& -\half \langle ij\rangle^2\eta^{\mu\nu} \label{anticom-}\\
\lbrack F^+_i,F^-_j \rbrack &=& 0 \label{com+-}\\
{\rm tr}\,(F_i^+F_j^-) &=& 0 \label{tr+-}
\ear
Moreover, as expected on general grounds
\cite{dufish,bardeen,selivanov,cangemi,chasie}
one finds the self duality properties

\bear
\tilde F^{\pm}_i &=& {\mp}iF^{\pm}_i
\label{sdrel}
\ear
With the help of these relations it is easy to compute the
two Maxwell invariants for the case of $F=F_{\rm tot}$:

\bear
\fourth F_{{\rm tot}\,\mn}\,F^{\mn}_{\rm tot}
&=&
\chi_+ + \chi_-
\non\\
\fourth F_{{\rm tot}\,\mn}\,\tilde F^{\mn}_{\rm tot}
&=&
-i(\chi_+ - \chi_-)
\non\\
\label{maxwelltochi}
\ear
where we have introduced

\bear
\chi_+ &\equiv & \half \sum_{1\le i < j \le N}[ij]^2 \non\\
\chi_- &\equiv & \half \sum_{1\le i < j \le N}\langle ij\rangle ^2 \non\\
\label{defchi}
\ear
Using (\ref{maxwelltochi}) in (\ref{relabFFtilde}) yields

\bear
a &=& \sqrt{\chi_+} + \sqrt{\chi_-} \non\\
b &=& -i(\sqrt{\chi_+}-\sqrt{\chi_-})  \non\\
\label{ab}
\ear
The choice of sign for $a,b$ does not matter since $a$ and $b$
appear only squared in the Lagrangian (\ref{L1spin}). Similarly,
there is no need to introduce a sign convention for
$\sqrt{\chi_{\pm}}$.

\noindent
Using (\ref{ab}) in (\ref{L1spin}) we get (omitting the subtraction terms)

\bear
{\cal L}_{\rm spin}^{(1)}\bigl(iF_{\rm tot}\bigr)
&=&
-
{1\over 8\pi^2}
\int_0^{\infty}{dT\over T}
\,\e^{-m^2T}
{(\sqrt{\chi_+}+\sqrt{\chi_-})
(\sqrt{\chi_+}-\sqrt{\chi_-})
\over \tan\bigl((\sqrt{\chi_+}+\sqrt{\chi_-})T\bigr)
\tan\bigl((\sqrt{\chi_+}-\sqrt{\chi_-})T\bigr)}
\non\\
\label{L1spinchi}
\ear
As we explained in the introduction, the right hand side constitutes
a generating functional for the one-loop (on-shell) photon amplitudes.
The $N$ - photon amplitude will be obtained by a double truncation of this
formal expression:
First, it  must be expanded in powers of  $\chi_+,\chi_-$, and only the part
of order $F_{\rm tot}^N$  kept from this series. Then, from the result those
terms should be extracted involving each individual $F_i$ just once.

Using the Taylor series,

\bear
{x\over \tan x} &=&
\sum_{n=0}^{\infty}(-1)^n{2^{2n}{\cal B}_{2n}\over (2n)!}x^{2n}
\label{taylorcot}
\ear
(the ${\cal B}_{2n}$ are Bernoulli numbers) the first step yields

\bear
{\cal L}_{\rm spin}^{(1)}\bigl(iF_{\rm tot}\bigr)
&=&
-{m^4\over 8\pi^2}
\sum_{N=4}^{\infty}
\Bigl({2e\over m^2}\Bigr)^N
\sum_{{K=0}\atop{K\,{\rm even}}}^{N}
\,c_{\rm spin}^{(1)}\Bigl({K\over 2},{N-K\over 2}\Bigr)
\chi_+^{K\over 2}\chi_-^{N-K\over 2}
\non\\
\label{L1spinchiexpand}
\ear
where

\bear
c_{\rm spin}^{(1)}\Bigl({K\over 2},{N-K\over 2}\Bigr)
&=&
(-1)^{N\over 2}(N-3)!
\sum_{k=0}^K\sum_{l=0}^{N-K}
(-1)^{N-K-l}
{{\cal B}_{k+l}{\cal B}_{N-k-l}
\over
k!l!(K-k)!(N-K-l)!}
\non\\
\label{defc1spin}
\ear
Here we have omitted the irrelevant terms of order
$\chi_{\pm}^0$, $\chi_{\pm}^1$. According to
the above, the amplitude with $K$ `$+$'
and $N-K$ `$-$' helicities is obtained from the
corresponding term in the sum (\ref{L1spinchiexpand})
by picking out the terms multilinear in the $F_i$'s.
It is immediately seen that
such terms exist only if $K$ is an even number.
Thus all amplitudes with an odd number of `$+$' helicities
do, in fact, vanish in the low energy limit.
For $K$ even, let us define

\bear
\chi_K^+ &\equiv & (\chi_+)^{K\over 2}
\Big\vert_{\rm all\,\, different}\non\\
&=&
{({\frac{K}{2}})!
\over 2^{K\over 2}}
\Bigl\lbrace
[12]^2[34]^2\cdots [(K-1)K]^2 + {\rm \,\, all \,\, permutations}
\Bigr\rbrace
\non\\
\chi_{N-K}^- &\equiv & (\chi_-)^{N-K\over 2}
\Big\vert_{\rm all\,\, different}\non\\
&=&
{({\frac{N-K}{2}})!
\over 2^{N-K\over 2}}
\Bigl\lbrace
\langle (K+1)(K+2)\rangle^2\langle (K+3)(K+4)\rangle^2\cdots
\langle (N-1)N\rangle^2 + {\rm \,\, all \,\, perm.}
\Bigr\rbrace
\non\\
\label{defchiKL+-}
\ear
The final result for the amplitude can then be written as

\bear
\Gamma_{\rm spin}^{(1)(EH)}
[\pol_1^+;\ldots ;\pol_K^+;\pol_{K+1}^-;\ldots ;\pol_N^-]
&=&
-{m^4\over 8\pi^2}
\Bigl({2e\over m^2}\Bigr)^N
\,c_{\rm spin}^{(1)}\Bigl({K\over 2},{N-K\over 2}\Bigr)
\chi_K^+\chi_{N-K}^-\non\\
\label{res1lspin}
\ear
(here and in the following we omit the momenta
$k_1,\ldots,k_N$ in the argument of amplitudes).

We remark that the introduction of the variables $\chi_{\pm}$ is not essential in this calculation.
An alternative,
though less elegant, way of arriving at the same result would be to expand
${\cal L}(iF_{\rm tot})$ directly in powers of $F_{\rm tot}$, perform the truncation to
the multilinear part of the order $F_{\rm tot}^N$ terms, and only after this use the spinor helicity
identities (\ref{anticom+}),(\ref{anticom-}).

\section{One-loop scalar QED}
\label{section1lscal}
\renewcommand{\theequation}{3.\arabic{equation}}
\setcounter{equation}{0}

The scalar QED case is completely analogous, and we will
write down only the main formulas. The analogue of the
Euler-Heisenberg Lagrangian (\ref{L1spin}) for the
scalar QED case was
given by Schwinger
\cite{schwinger51}:

\bear
{\cal L}_{\rm scal}^{(1)}
&=&
{1\over 16\pi^2}
\int_0^{\infty}{dT\over T}
\,\e^{-m^2T}
\biggl\lbrack
{e^2ab\over \sinh(eaT)\sin(ebT)}
+{e^2\over 6} (a^2-b^2) -{1\over T^2}
\biggr]
\non\\
\label{L1scal}
\ear
Using the Taylor expansion

\bear
{x\over\sin x}&=&
-\sum_{n=0}^{\infty}
(-1)^n{\bigl(2^{2n}-2\bigr){\cal B}_{2n}
\over (2n)!}x^{2n}
\label{taylorcosec}
\ear
we can use the same procedure as in the spinor QED case.
Again the result vanishes for
odd $K$, and for even $K$ one obtains
a formula analogous to (\ref{res1lspin}):

\bear
\Gamma_{\rm scal}^{(1)(EH)}
[\varepsilon_1^+;\ldots ;\pol_K^+;\pol_{K+1}^-;\ldots ;\pol_N^-]
&=&
{m^4\over 16\pi^2}
\Bigl({2e\over m^2}\Bigr)^N
\,c_{\rm scal}^{(1)}\Bigl({K\over 2},{N-K\over 2}\Bigr)
\chi_K^+\chi_{N-K}^-\non\\
\label{res1lscal}
\ear
where now

\bear
c_{\rm scal}^{(1)}\Bigl({K\over 2},{N-K\over 2}\Bigr)
&=&
(-1)^{N\over 2}(N-3)!
\sum_{k=0}^K\sum_{l=0}^{N-K}
(-1)^{N-K-l}
\non\\
&&\times
{
\bigl(1-2^{1-k-l)}\bigr)
\bigl(1-2^{1-N+k+l}\bigr)
{\cal B}_{k+l}{\cal B}_{N-k-l}
\over
k!l!(K-k)!(N-K-l)!}
\non\\
\label{defc1scal}
\ear

\section{Two-loop scalar and spinor QED}
\label{section2l}
\renewcommand{\theequation}{4.\arabic{equation}}
\setcounter{equation}{0}

The following integral representation was obtained
in \cite{ritusspin} for the two-loop generalization of the
Euler-Heisenberg Lagrangian (\ref{L1spin}):

\bear
{{\cal L}_{\rm spin}^{(2)}}
&=&
{\alpha\over 16\pi^3}
\int_0^{\infty}dT
\Biggl\lbrace
\int_0^TdT'
\biggl\lbrack
K(T,T') - {K_0(T)\over T'}
\biggr\rbrack
+K_0(T)\Bigl({\rm ln}(m^2T)+\gamma -{5\over 6}\Bigr)
\Biggr\rbrace
\non\\
\label{L2spin}
\ear
where

\bear
K(T,T') &=&
\e^{-m^2 (T+T')}
\Biggl\lbrace
{a^2b^2\over PP'}
\Bigl[4m^2(SS'+PP')I_0+I\Bigr]
\non\\&&\hspace{-30pt}
-{1\over TT'(T+T')}
\biggl[
4m^2 +{2\over T+T'} + {a^2-b^2\over 3}
\Bigl(2m^2(2T^2+2T'^2-TT')-{5TT'\over T+T'}\Bigr)
\biggr]
\Biggr\rbrace\non\\
K_0(T) &=& \e^{-m^2T}
\Bigl(4m^2 -{\partial\over \partial T}\Bigr)
\biggl(
{ab\over\tanh (aT)\tan (bT)}-{1\over T^2} -{a^2-b^2\over 3}
\biggr) \non\\
\label{defKspins}
\ear

\bear
I_0 &=& {1\over B-A}\ln ({B\over A})\non\\
I&=& {q-p\over B-A}I_0-{{q\over B}-{p\over A}\over B-A}\non\\
p&=& 2{a^2\cos (b(T'-T))\over \sinh(aT)\sinh(aT')},\quad
q= 2{b^2\cosh (a(T'-T))\over\sin(bT)\sin(bT')}\non\\
P&=& {\rm sinh}(aT)\sin (bT),\qquad\quad
S= \cosh (aT)\cos (bT)\non\\
A&=& a\bigl(\coth (aT) + \coth (aT')\bigr),\quad
B= b\bigl(\cot (bT) + \cot (bT')\bigr)\non\\
\label{defspin}
\ear
Here $\gamma $ is the Euler-Mascheroni constant. The charge $e$ has been
set to unity.

In contrast to the one-loop formula (\ref{L1spin})
it is not known how to obtain from this
integral representation a closed-form expression
for the coefficients of the weak field expansion.
Therefore, at two loops we contend ourselves
with a calculation of this expansion to a certain order.
Using MATHEMATICA we have found it straightforward to
compute this expansion up to the order $(F^{10})$.
As in the one-loop case, from the
resulting polynomial in $a,b$ we can directly read off
the helicity amplitudes for $N=4,6,8,10$. To obtain a nonvanishing result,
again we have to assume that not only $N$ but also $K$
are even. Its form is analogous to (\ref{res1lspin}):

\bear
\Gamma_{\rm spin}^{(2)(EH)}
[\varepsilon_1^+;\ldots \pol_K^+;\pol_{K+1}^-;\ldots ;\pol_N^-]
&=&
-{\alpha \pi m^4\over 8\pi^2}
\Bigl({2e\over m^2}\Bigr)^N
\,c_{\rm spin}^{(2)}\Bigl({K\over 2},{N-K\over 2}\Bigr)
\chi_K^+\chi_{N-K}^-\non\\
\label{res2lspin}
\ear
where the coefficients $c_{\rm spin}^{(2)}\bigl({K\over 2},{N-K\over 2}\bigr)$
are given in the appendix.

\vspace{20pt}
\noindent
For the scalar QED case, we use the similar representation \cite{ritusscal}

\bear
{{\cal L}_{\rm scal}^{(2)}}
&=&
-{\alpha\over 32\pi^3}
\int_0^{\infty}dT
\Biggl\lbrace
\int_0^TdT'
\biggl\lbrack
\tilde K(T,T') - {\tilde K_0(T)\over T'}
\biggr\rbrack
+\tilde K_0(T)\Bigl({\rm ln}(m^2T)+\gamma -{7\over 6}\Bigr)
\Biggr\rbrace
\non\\
\label{ritus2lscal}
\ear
where now
\vfill\eject

\bear
\tilde K(T,T') &=&
\e^{-m^2(T+T')}
\Biggl\lbrace
{a^2b^2\over PP'}
\Bigl[m^2I_0-{\tilde I\over 2}\Bigr]
\non\\&&\hspace{-40pt}
-{1\over TT'(T+T')}
\biggl[
m^2-{1\over T+T'} - {a^2-b^2\over 6}
\Bigl(m^2(T+T')^2-m^2TT'-{11TT'\over (T+T')  }\Bigr)
\biggr]
\Biggr\rbrace\non\\
\tilde K_0(T) &=& \e^{-m^2T}
\Bigl(m^2 +\half {\partial\over \partial T}\Bigr)
\biggl(
{ab\over\sinh (aT)\sin (bT)}-{1\over T^2} + {a^2-b^2\over 6}
\biggr) \non\\
\label{Kscal}
\ear

\bear
\tilde I&=& {\tilde q-\tilde p\over B-A}
I_0-{{\tilde q\over B}-{\tilde p\over A}\over B-A}\non\\
\tilde p&=& 2a^2\Bigl({\rm coth}(aT){\rm coth}(aT')-3\Bigr),\qquad
\tilde q = 2b^2\Bigl({\rm cot}(bT){\rm cot}(bT')+3\Bigr)\non\\
\label{defscal}
\ear
$I_0,P,A,B$ are as in (\ref{defspin}).
Computation of the weak field expansion to the same order
$(F^{10})$ yields

\bear
\Gamma_{\rm scal}^{(2)(EH)}
[\varepsilon_1^+;\ldots \pol_K^+;\pol_{K+1}^-;\ldots ;\pol_N^-]
&=&
{\alpha\pi m^4\over 16\pi^2}
\Bigl({2e\over m^2}\Bigr)^N
\,c_{\rm scal}^{(2)}\Bigl({K\over 2},{N-K\over 2}\Bigr)
\chi_K^+\chi_{N-K}^-\non\\
\label{res2lscal}
\ear
with coefficients also given in the appendix.

\section{Conclusions}
\label{conclusions}
\renewcommand{\theequation}{5.\arabic{equation}}
\setcounter{equation}{0}
To summarize, we have shown here that the use of the
effective action, when combined with spinor helicity
techniques, provides a simple and elegant way
to obtain information on the low energy limit of the
QED $N$ photon amplitudes. This has allowed us to
derive an explicit
formula for the one-loop $N$ point amplitudes, as well
as for the two-loop amplitudes up to the ten-point case.
In particular, it has turned out that all amplitudes
with an odd number of `+' helicities vanish in the low energy
limit. From the approach presented here it is clear that
this property follows directly from the fact that the constant field
effective action can be written as a function of the two
Maxwell invariants. We therefor conclude that this vanishing must
persist to all loop orders. Since these amplitudes are not forbidden
by any known symmetries, and indeed,
the one-loop four-point $(+++-)$ amplitude is known to be non-vanishing
with full momentum \cite{cotopi},
this comes rather unexpected (for the special case of the amplitudes
with all helicities but one positive this vanishing
had been noted
already in \cite{dunsch3}).

Obviously, the self-duality relations
fulfilled by field strength tensors with definite
helicities, eqs. (\ref{sdrel}) play an important part in these simplifications.
We expect that these relations, as well as the
variables $\chi_{\pm}$, will also have a useful role
to play for the photon amplitudes at full momentum.
One indication for this is the appearance of factors
of traces of products of field strength tensors
in the parameter integrals for the $N$ - photon
amplitudes generated by the Bern-Kosower formalism
\cite{berkos,berdun,strassler1,strassler2,ss1,nphoton}.
Work in this direction is in progress.

\vspace{10pt}
\noindent
{\bf Acknowledgements:}
We would like to thank L. Dixon for information concerning
\cite{dixon}. L. Martin thanks the IFM, UMSNH, for their hospitality and
the Secretaria de Relaciones Exteriores for funding.
We are also indebted to the referee for several important remarks.

\vfill\eject

\section{Appendix: Two-loop coefficients}
\label{appendix}
\renewcommand{\theequation}{A.\arabic{equation}}
\setcounter{equation}{0}

\def\arraystretch{2.5}
\nopagebreak[3]
\nopagebreak[3]
\begin{centering}
\nopagebreak[3]
\vspace{20pt}
\begin{tabular}{|l||c|c|}
\hline
$c^{(2)}(\frac{K}{2},\frac{N-K}{2})$  & Scalar QED & Spinor QED \cr\hline \hline
$c^{(2)}(5,0)$& $ \frac{611}{80640 \pi^2}     $& $\frac{317}{40320 \pi^2}$\cr\hline
$c^{(2)}(4,1)$& $\frac{349609}{3628800 \pi^2} $& $\frac{-8707}{1814400 \pi^2}$\cr\hline
$c^{(2)}(3,2)$& $\frac{688637}{2332800\pi^2}  $& $\frac{-3190547}{8164800 \pi^2}$\cr\hline \hline
$c^{(2)}(4,0)$& $\frac{67}{12800 \pi^2}       $& $\frac{2221}{403200\pi^2}$\cr\hline
$c^{(2)}(3,1)$& $\frac{273619}{6350400 \pi^2} $& $\frac{-151379}{6350400 \pi^2}$\cr\hline
$c^{(2)}(2,2)$& $\frac{2055163}{25401600\pi^2}$& $\frac{-37763}{282240 \pi^2}$\cr\hline \hline
$c^{(2)}(3,0)$& $\frac{13}{1920 \pi^2}        $& $\frac{7}{960 \pi^2}$\cr\hline
$c^{(2)}(2,1)$& $\frac{8563}{259200 \pi^2}     $& $\frac{-5821}{129600 \pi^2}$\cr\hline \hline
$c^{(2)}(2,0)$& $\frac{3}{128 \pi^2}          $& $\frac{5}{192 \pi^2}$\cr\hline
$c^{(2)}(1,1)$& $\frac{307}{5184 \pi^2}       $& $\frac{-391}{2592 \pi^2}$\cr\hline
\end{tabular}
\vspace{20pt}
\label{t1}
\end{centering}

\vfill\eject

\end{document}